\documentclass[prc,aps,groupedaddress,superscriptaddress,twocolumn,nofootinbib,showpacs,showkeys,floatfix,a4paper,10pt,amsmath,amssymb]{revtex4-1}
\usepackage{graphicx}
\usepackage[english]{babel}
\usepackage{dcolumn}
\usepackage{bm}

\begin{document}

\title{Beyond-mean-field description of octupolarity in dysprosium isotopes with 
the Gogny-D1M energy density functional}

\author{R. Rodr\'{\i}guez-Guzm\'an}
\email{raynerrobertorodriguez@gmail.com}
\affiliation{Departamento de F\'isica Aplicada I, Escuela Polit\'ecnica Superior, Universidad de Sevilla, Seville, E-41011, Spain}

\author{L.M. Robledo}
\email{luis.robledo@uam.es}
\affiliation{%
Center for Computational Simulation, Universidad Polit\'ecnica de 
Madrid, Campus Montegancedo, 28660 Boadilla del Monte, Madrid, Spain
}%
\affiliation{Departamento  de F\'{\i}sica Te\'orica and CIAFF, 
Universidad Aut\'onoma de Madrid, 28049-Madrid, Spain}

\date{\today}

\begin{abstract} 
The emergence and  stability of (static) octupole deformation effects 
in Dy isotopes from dripline to dripline ($72 \le N \le 142$) is analyzed in this paper 
using mean-field and beyond-mean-field techniques often used for this 
purpose. We find  static octupole deformations at the 
Hartree-Fock-Bogoliubov (HFB) level with the Gogny D1M force for $N 
\approx 134$ isotopes, while nuclei with $N \approx 88$ exhibit 
reflection-symmetric ground states. It is shown that, given the 
softness found in the mean-field and parity-projected potential energy 
surfaces along the octupole direction, neither of these two levels of 
approximation is suficcient to extract conclusions about the (permanent 
and/or vibrational) nature of octupole dynamic in Dy isotopes. 
From the analysis of the collective wave functions as well 
as the excitation energies of the first negative-parity states and 
$B(E3)$ strengths, obtained within the framework of a two-dimensional 
symmetry-conserving generator coordinate method (2D-GCM), it is 
concluded that the increased octupole collectivity in Dy isotopes  with 
$N \approx 88$ and $N \approx 134$ is a vibrational-like effect that is 
not directly related to permanent mean-field octupole deformation in 
the considered nuclei. A pronounced suppression of the $B(E1)$ strengths  
is predicted for isotopes with  $N \approx 82$ and  $N \approx 126$. 
The comparison of results obtained with other parametrizations, show the robustness 
of the predicted trends with respect to the underlying Gogny energy 
density functional.
\end{abstract}


\maketitle{}

%
%
%

\section{Introduction.}

The majority of the spherical and/or quadrupole-deformed nuclear ground 
states are reflection-symmetric. However, due to the mean-field 
spontaneous symmetry breaking mechanism \cite{rs} reflection-asymmetric 
ground states tend to be favored energetically in certain regions of 
the nuclear chart \cite{butler_96}. Those regions are usually associated with the 
neutron/proton numbers $N/Z = 34, 56, 88$ and $134$ where the coupling 
between  intruder $(N+1,l+3,j+3)$ and normal-parity $(N,l,j)$  states 
is more effective in developing octupole deformed ground states. 
Octupole-related features have been studied around the already 
mentioned neutron/proton numbers, however, the search for new islands 
of reflection-asymmetric shapes, all over the  nuclear chart, still 
represents one of the frontiers in nuclear structure physics nowadays. 
Within this context, a better understanding of the permanent and/or 
vibrational nature of octupole dynamic in atomic nuclei still represents a 
major challenge that cannot be resolved with plain mean field 
calculations.

Octupolarity along the Dy isotopic chain has been the subject of 
experimental studies. For example, bands associated with parity 
doublets have been studied in $^{157}$Dy using the JUROGAM II array 
\cite{spect-157Dy}. A rotational band, built on an octupole vibration, 
has been identified in $^{152}$Dy \cite{exp-152Dy}. The $E1$ 
transitions between opposite parity bands, have been studied in  
$^{154}$Dy \cite{exp-154Dy}. Negative parity bands have also been 
investigated in both $^{156}$Dy and $^{162}$Dy 
\cite{exp-156Dy,exp-162Dy}. The experimental findings 
\cite{spect-157Dy,exp-152Dy,exp-154Dy,exp-156Dy,exp-162Dy} raise 
questions about the impact of octupole correlations in the structural evolution 
along the Dy isotopic chain as well as on the (permanent and/or 
vibrational) nature of octupole deformation effects in those isotopes. 
Recently, relativistic mean-field calculations have been carried out 
for Dy nuclei \cite{microscopic-22}. On the basis of plain mean-field 
results, it has been concluded, that $N \approx 88$ and $N \approx 134$ Dy isotopes 
exhibit permanent octupole deformation. The conclusion extends to isotopes where the octupole minima 
found in the calculations are very shallow and the corresponding 
potential energy surfaces exhibit a rather soft behavior along the 
octupole direction. The conclusions of Ref.~\cite{microscopic-22} are 
at variance with previous macroscopic-microscopic (Mac-Mic) results 
\cite{MM-6} as well as with the ones extracted from this microscopic 
study in which, the relevance of beyond-mean-field octupole dynamics in 
Dy isotopes is considered with the Gogny energy density functional 
(EDF) \cite{gogny}, using the models already introduced in 
Refs.~\cite{q2q3-rayner-1,q2q3-rayner-2,q2q3-rayner-3,q2q3-rayner-4} 
and used to describe octupole dynamics in other regions of the nuclear 
chart. In particular, we address in the present study the stability of 
(static) mean-field octupole deformation effects once beyond-mean-field 
symmetry restoration and/or configuration mixing (dynamical) effects 
are taken into account. To this end, calculations have been carried out 
along the Dy isotopic chain from proton to neutron dripline ($72 \le N \le 142$).

A lot of effort has  been devoted to  better understand basic 
fingerprints of octupole correlations (see, for example, 
Refs.~\cite{Ahmad_93,butler_2016,butler_2015,Tandel-oct,Li-oct,Ahmad-oct-2,Bucher-oct-1,Bucher-oct-2,Bernard16,Butler_2020,Gaffney-2013} 
and references therein).  Previous experiments have found  evidence for 
octupole deformed ground states in $^{144,146}$Ba 
\cite{Bucher-oct-1,Bucher-oct-2} and  $^{222,224}$Ra 
\cite{Butler_2020,Gaffney-2013}. The measured low-lying states in 
$^{224,226}$Rn suggest that those isotopes should be characterized as octupole vibrations 
\cite{Butler-2019-Rn}. Furthermore, fingerprints of octupole 
correlations have also been found in the case of $^{228}$Ra and 
$^{228}$Th \cite{Butler_2020,Chishti-2020}. Here, one should  keep in 
mind that renewed interest in octupole correlations  also comes from 
the need to improve the description of fission paths in heavy and 
super-heavy nuclei. In particular octupole correlations are well known 
to affect the outer sectors of the fission paths in those nuclei (see, 
for example, 
Refs.~\cite{fission-exampleq3-1,fission-exampleq3-2,fission-exampleq3-3,fission-exampleq3-4} 
and references therein). Octupole deformation is also one of the 
collective coordinates at play in the case of cluster radioactivity 
\cite{cluster}.
 
A wide range of models has been employed to study octupole dynamics. 
For example, octupole shapes have been considered within the 
Mac-Mic framework 
\cite{MM-1,MM-2,MM-3,MM-4,MM-5,MM-6} as well as within the mapped 
Interacting Boson Model (IBM) 
\cite{mapped-IBM-1,mapped-IBM-2,mapped-IBM-3,mapped-IBM-4,mapped-IBM-5,mapped-IBM-6,mapped-IBM-7,mapped-IBM-8}. 
Octupole correlations have also been the subject of intense microscopic 
scrutiny, both at the mean-field level and beyond, using  
non-relativistic and relativistic approximations 
\cite{microscopic-1,microscopic-2,microscopic-3,microscopic-4,microscopic-5,microscopic-6,microscopic-7,microscopic-8,microscopic-10,microscopic-11,microscopic-12,microscopic-13,microscopic-14,microscopic-15,microscopic-16,microscopic-17,microscopic-18,microscopic-19,microscopic-20,microscopic-21}.

In the case of the Gogny energy density functional (EDF) \cite{gogny}, 
the models of 
Refs.~\cite{q2q3-rayner-1,q2q3-rayner-2,q2q3-rayner-3,q2q3-rayner-4} 
have already been employed to study the quadrupole-octupole coupling in 
regions of the nuclear chart such as the Sm and Gd isotopes with $84 
\le N \le 92$ \cite{q2q3-rayner-1}, actinide nuclei with neutron number 
$N \approx 134$ \cite{q2q3-rayner-2}, Rn, Ra and Th isotopes 
\cite{q2q3-rayner-3} as well as neutron-rich actinides and super-heavy 
nuclei \cite{q2q3-rayner-4}. First, the quadrupole $Q_{20}$ and 
octupole $Q_{30}$ deformation parameters have been considered 
simultaneously within the constrained Hartree-Fock-Bogoliubov (HFB) 
framework \cite{rs} to build the corresponding $(Q_{20},Q_{30})$ 
mean-field potential energy surfaces (MFPESs). Second, the changes 
induced in the MFPESs by the restoration of the reflection symmetry 
have been considered by projecting the $(Q_{20},Q_{30})$-constrained 
intrinsic HFB states onto a good parity. Third, the quadrupole-octupole 
coupling has been taken into account using a two-dimensional 
symmetry-conserving Generator Coordinate Method (2D-GCM) ansatz
\cite{q2q3-rayner-1,q2q3-rayner-2,q2q3-rayner-3,q2q3-rayner-4}.

\begin{figure*}
\includegraphics[width=0.95\textwidth]{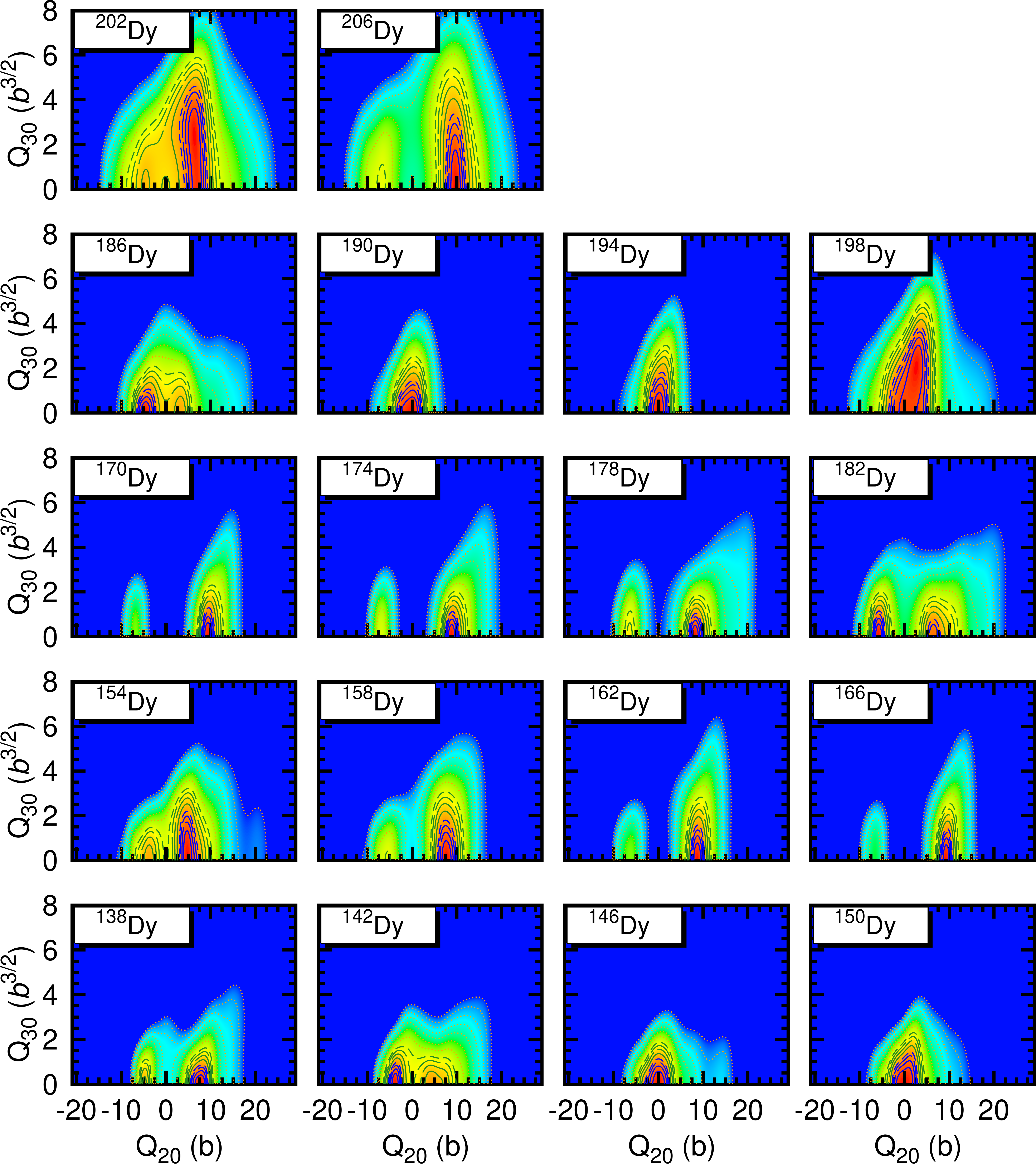}
\caption{(Color online) 
MFPESs computed with the Gogny-D1M EDF for a selected set of Dy 
isotopes. Dark blue contour lines extend from 0.25 MeV up to 1 MeV 
above the ground state energy in steps of 0.25 MeV in the ascending 
sequence full, long-dashed, medium-dashed and short-dashed. Dark green 
contour lines extend from  1.5 MeV up to 3 MeV above the ground state 
in steps of 0.5 MeV with the same sequence of full, long-dashed, 
medium-dashed and short-dashed as before. From there on, orange dotted 
contour lines are drawn in steps of 1 MeV. The color code spans a range 
of 10 MeV with red corresponding to the lowest energy and blue 
corresponding to an energy 10 MeV above. The intrinsic HFB energies are 
symmetric under the exchange $Q_{30} \rightarrow -Q_{30}$. For $A = 
150$ a quadrupole deformation $Q_{20}= 5 b$ is equivalent to $\beta_{2} 
= 0.217$, whereas  for $A = 200$ an octupole deformation $Q_{30}= 2.5 
b^{3/2}$ is equivalent to $\beta_{3} = 0.113$. For more details, see 
the main text.
}
\label{mean-field-surfaces} 
\end{figure*}

The key lesson extracted from the studies mentioned above 
\cite{q2q3-rayner-1,q2q3-rayner-2,q2q3-rayner-3,q2q3-rayner-4} is that, 
for the considered nuclei, 2D-GCM zero-point quantum fluctuations are 
essential to obtain a systematic of the $B(E1)$ and $B(E3)$ strengths 
as well as of  the excitation energies of the lowest negative-parity 
states that accounts reasonably well for the available experimental 
data. Moreover, it has also been shown that such 2D-GCM quantum fluctuations 
can lead to an enhanced octupolarity 
as well as to a weaker dependence of the correlation energies with 
neutron number. In this respect, we also refer the reader to previous 
large scale surveys, using the octupole degree of freedom as a single 
generating coordinate \cite{microscopic-20,microscopic-21}.

The main aim of this paper is to address, the stability of mean-field 
octupole deformation effects as well as the impact of beyond-mean-field 
(dynamical) correlations in dripline-to-dripline calculations for Dy 
isotopes. Our results reexamine the conclusions of relativistic 
mean-field \cite{microscopic-22} studies around both $N=88$ and 
$N=134$ pointing to permanent octupole deformation effects in those 
regions. In order to disentangle the role of static octupole 
deformation, we have first obtained a set of 
$(Q_{20},Q_{30})$-constrained Gogny-HFB wave functions for the 
even-even isotopes $^{138-208}$Dy. The energies corresponding to each 
of these mean-field states are then used to build the MFPESs as 
functions of the quadrupole $Q_{20}$ and octupole $Q_{30}$ 
deformations. Note, that the considered range of neutron numbers, i.e., 
$72 \le N \le 142$, includes the octupole magic number $N=88$ and 
extends up to a very neutron-rich sector to also include the octupole 
magic number $N=134$. Therefore, the Gogny-HFB calculations allow us to 
examine the emergence and evolution of static ground state 
reflection-asymmetric shapes along the Dy isotopic chain and, in 
particular, to compare with Mac-Mic \cite{MM-6} and relativistic 
mean-field \cite{microscopic-22} predictions around both $N=88$ and 
$N=134$.

As will be shown later on in the paper, for the studied isotopes, the 
MFPESs often are rather soft along the $Q_{30}$-direction and/or the 
mean-field octupole correlation energies $E_{CORR,HFB}$ [see, 
Eq.(\ref{oct-def-ener})], are rather small. Moreover, in some cases the 
MFPESs exhibit a transitional behavior along the $Q_{20}$-direction. 
Taking into account the experience, obtained in previous works 
\cite{q2q3-rayner-1,q2q3-rayner-2,q2q3-rayner-3,q2q3-rayner-4}, on the 
role of dynamical correlations in such scenarios and the mean-field 
results already mentioned, we have then studied the impact of 
beyond-mean-field zero-point quantum fluctuations in $^{138-208}$Dy. To 
this end, we have resorted to both parity symmetry restoration and 
symmetry-conserving 2D-GCM quadrupole-octupole configuration mixing
\cite{q2q3-rayner-1,q2q3-rayner-2,q2q3-rayner-3,q2q3-rayner-4}.

The results discussed in this paper, at the three levels of 
approximation employed, have been obtained with the parametrization D1M 
\cite{gogny-d1m} of the Gogny-EDF. The parametrization D1M has already 
been shown to provide a reasonable description of octupole-related 
features in previous studies 
\cite{q2q3-rayner-1,q2q3-rayner-2,q2q3-rayner-3,q2q3-rayner-4}. 
However, in some instances, we will also discuss results obtained with 
the parametrizations D1S \cite{gogny-d1s} and D1M$^{*}$ 
\cite{gogny-d1mstar} in order to illustrate the robustness of the 
predictions with respect to the underlying Gogny-EDF.

The paper is organized as follows. The HFB and beyond-mean-field 
approximations employed in this study are briefly outlined in 
Secs.~\ref{mf-results} and \ref{beyond-mf-results}. The results 
obtained with the corresponding approach will be discussed in each 
section. The HFB results will be discussed in Sec.~\ref{mf-results}, 
while dynamical beyond-mean-field  correlations are considered in 
Sec.~\ref{beyond-mf-results}. In particular, parity symmetry 
restoration is considered in Sec.~\ref{parity-proj-results}, while 
symmetry-conserving 2D-GCM quadrupole-octupole configuration mixing is 
discussed in Sec.~\ref{2D-GCM-results}. In this 
Sec.~\ref{2D-GCM-results}, the excitation energies of the lowest 
negative-parity states as well as $B(E1)$ and $B(E3)$ strengths 
obtained for $^{138-208}$Dy will be discussed and compared with the 
available experimental data \cite{EXP-E3-Be-Kibedi}. Furthermore, we 
will also illustrate the robustness of the 2D-GCM predictions with 
respect to the underlying Gogny-EDF. Finally, Sec.~\ref{conclusions} is 
devoted to the concluding remarks.

%
%
%

\section{Results}
\label{results}

In this work we study the emergence and stability of octupole 
deformation effects in the isotopic chain $^{138-208}$Dy from a 
microscopic point of view using the  density-dependent Gogny-D1M EDF. 
To this end, the HFB approach \cite{rs}, with constrains on the axially 
symmetry quadrupole $\hat{Q}_{20}$ and octupole $\hat{Q}_{30}$ 
operators, is employed as a first step. On the other hand dynamical 
beyond-mean-field correlations  are considered via parity projection of 
the intrinsic HFB states and/or symmetry-conserving 2D-GCM 
quadrupole-octupole configuration mixing. In this section, we briefly 
outline  these approaches  
\cite{q2q3-rayner-1,q2q3-rayner-2,q2q3-rayner-3,q2q3-rayner-4} and 
discuss the results obtained with each of them.


\subsection{Hartree-Fock-Bogoliubov}
\label{mf-results}


We have first performed 
$(Q_{20},Q_{30})$-constrained Gogny-D1M HFB calculations for $^{138-208}$Dy. In the
calculations the HFB equation has been solved with constrains 
on the axially symmetric quadrupole 
\begin{equation}
\hat{Q}_{20} = \frac{1}{2} \Big(x^{2} + y^{2} \Big)
\end{equation}
and octupole
\begin{equation}
\hat{Q}_{30} = z^{2}- \frac{3}{2} \Big(x^{2} + y^{2} \Big)z
\end{equation}
operators, using an approximate second-order gradient method 
\cite{second-order-grad}. A constrain on the operator
$\hat{Q}_{10}$ has also been used to fix the center of mass at
the origin
\cite{q2q3-rayner-1}. The HFB quasiparticle operators \cite{rs}
have been expanded in a  (deformed) axially
symmetric harmonic oscillator (HO) basis  containing 
15 major shells. Axial symmetry has been kept as a 
self-consistent symmetry in order to alleviate the 
computational effort. 

For each of the intrinsic states
$ |\Phi (Q_{20},Q_{30}) \rangle$ obtained in the constrained Gogny-HFB calculations, the quadrupole 
$Q_{20}$
and octupole $Q_{30}$ deformations are defined as the mean values
\begin{equation}
Q_{20} = \langle \Phi (Q_{20},Q_{30}) | \hat{Q}_{20} | \Phi(Q_{20},Q_{30}) \rangle
\end{equation}
and
\begin{equation}
Q_{30} = \langle \Phi (Q_{20},Q_{30}) | \hat{Q}_{30} | \Phi(Q_{20},Q_{30}) \rangle
\end{equation}
The corresponding deformation parameters $\beta_{\lambda}$ ($\lambda =2,3$) are then defined as 
\begin{equation}
\beta_{\lambda} = \frac{\sqrt{4 \pi (2\lambda +1)}}{3 R_{0}^{\lambda} A} Q_{\lambda 0}
\end{equation}
with $R_{0}=1.2A^{1/3}$ and A the mass number. For example, for
$A = 150$ a quadrupole deformation $Q_{20}= 5 b$ is equivalent to
$\beta_{2} = 0.217$, whereas  for $A = 200$ an octupole
deformation $Q_{30}= 2.5 b^{3/2}$ is equivalent to 
$\beta_{3} = 0.113$.

The Gogny-HFB MFPESs  
are depicted in Fig.~\ref{mean-field-surfaces} 
for a selected set of Dy isotopes,
as illustrative examples.  Those
MFPESs are nothing else than the HFB energies 
\begin{equation}
\label{HFB-energies}
E_{HFB}(Q_{20},Q_{30}) = 
\frac{\langle \Phi (Q_{20},Q_{30}) | \hat{H} | \Phi(Q_{20},Q_{30}) \rangle} 
{\langle \Phi (Q_{20},Q_{30}) |  \Phi(Q_{20},Q_{30}) \rangle}
\end{equation}
corresponding 
to each of 
the intrinsic states $ |\Phi (Q_{20},Q_{30}) \rangle$. 
The HFB energies (\ref{HFB-energies}) are invariant under the exchange
of $Q_{30}$ into $-Q_{30}$ to be associated to the parity symmetry of the
interaction
\begin{equation} 
E_{HFB}(Q_{20},Q_{30})= E_{HFB}(Q_{20},-Q_{30})
\end{equation}
As a consequence of this invariance, only the energies corresponding to
$Q_{30} \ge 0$ values are included in Fig.~\ref{mean-field-surfaces}.
In the calculations, the $Q_{20}$-grid $-25~b \le Q_{20} \le 35~b$ (with a step
$\delta Q_{20} =1~b$) and the $Q_{30}$-grid $ 0~b^{3/2} \le Q_{30} \le 10~b^{3/2}$
(with a step $\delta Q_{30} =~0.25~b^{3/2}$) have been employed.

\begin{figure}
\includegraphics[width=0.45\textwidth]{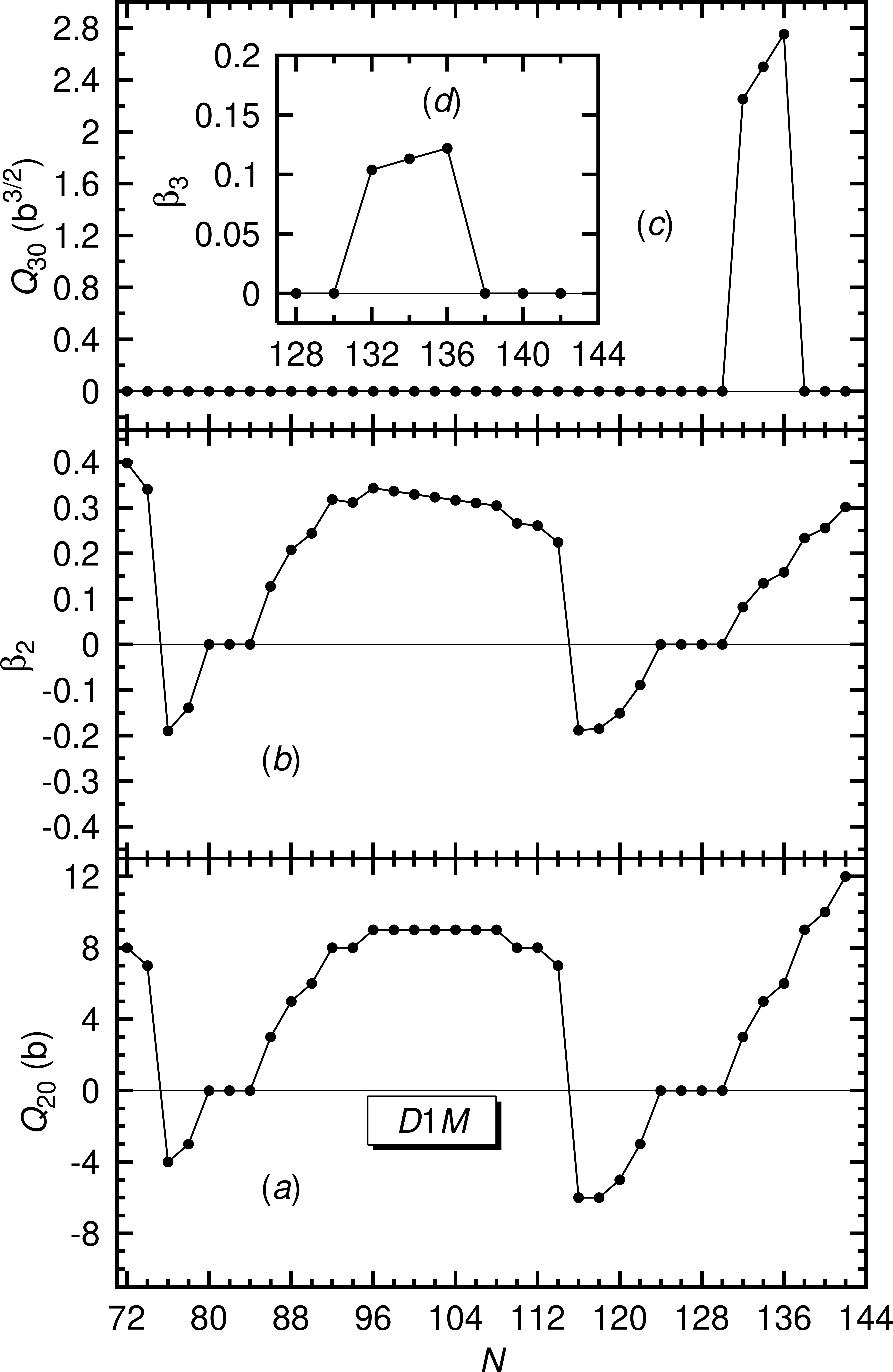}
\caption{The mean field ground state quadrupole (octupole)
deformations are plotted, as functions of the neutron number $N$, in panel
(a) [(c)] for $^{138-208}$Dy. The corresponding quadrupole (octupole) deformation parameters
$\beta_{2}$ ($\beta_{3}$) are also depicted in panel (b) [(d)]. Results 
have been obtained with the Gogny-D1M EDF.
}
\label{summary-def-mf} 
\end{figure}

The ground state quadrupole deformations are plotted, as functions of 
the neutron number $N$, in panel (a) of Fig.~\ref{summary-def-mf} for 
$^{138-208}$Dy. The corresponding $\beta_{2}$ deformation parameters 
are depicted in panel (b) of the same figure. Along the 
$Q_{20}$-direction there is a shape/phase transition from a prolate 
($^{138,140}$Dy) to an oblate ($^{142,144}$Dy) ground state, followed 
by spherical ground states in $^{146-150}$Dy, reflecting the proximity 
to the neutron shell closure $N=82$. With increasing neutron number, 
the ground state quadrupole deformations increase reaching  values of 
$Q_{20}~=8-9~b$  for $92 \le N \le 112$. This is, once more, followed 
by shape/phase transitions to oblate ground states in $^{182-188}$Dy 
and then to spherical ground states in $^{190-196}$Dy, associated with 
the proximity to the neutron shell closure $N=126$. For larger neutron 
numbers, the ground state quadrupole deformations exhibit a pronounced 
increase, reaching the value $Q_{20}~=12~b$ for $^{208}$Dy.

For the considered Dy isotopes, the ground state quadrupole 
deformations are within the range $-6~b \le Q_{20} \le 12~b$ ($-0.19 
\le \beta_{2} \le 0.34$). The results obtained with Gogny-D1M, as well 
as the ones obtained with the D1S and D1M$^{*}$ parametrizations, for 
the ground state quadrupole deformations agree well with previous 
Mac-Mic  \cite{MM-6} and reflection-asymmetric relativistic mean-field   
\cite{microscopic-22}  results. Note that, for some of the considered 
isotopes, the MFPESs depicted in Fig.~\ref{mean-field-surfaces} exhibit 
transitional features along the $Q_{20}$-direction.

As can be seen from 
Fig.~\ref{mean-field-surfaces}  and 
from panels (c) and (d)  of Fig.\ref{summary-def-mf}, static Gogny-D1M
ground state octupole deformations are only predicted 
for  $^{198-202}$Dy, i.e., for very neutron-rich 
isotopes around $N=134$. In this case, the 
ground state octupole deformations are within the range 
$2.25~b^{3/2} \le Q_{30} \le 2.75~b^{3/2}$ ($0.10 \le \beta_{3} \le 0.12$).
Octupole-deformed neutron-rich nuclei 
have already  been predicted, in this \cite{MM-6,microscopic-22} and 
other regions of the 
nuclear chart
\cite{q2q3-rayner-4,MM-4,microscopic-17,microscopic-18,microscopic-19,microscopic-5}.
The soft
behavior of the Gogny-D1M MFPESs 
along the  $Q_{30}$-direction, as one approaches the neutron number
$N=134$, becomes apparent from  
Fig.~\ref{mean-field-surfaces}. Nevertheless, even in the 
case of nuclei with octupole deformed mean-field
ground states (i.e., $^{198-202}$Dy), the
HFB energy gained by breaking reflection symmetry 
\begin{equation}
\label{oct-def-ener}
E_{CORR,HFB} = E_{HFB,Q30 =0} - E_{HFB,GS}
\end{equation}
and defined as
the difference between the HFB energy corresponding to the absolute minimum
obtained in reflection-symmetric calculations and the 
energy corresponding to the absolute minimum
of the  $(Q_{20},Q_{30})$-MFPES, is rather small 
(188, 266 and 70 $KeV$ for $^{198-202}$Dy, respectively).

The MFPESs shown in Fig.~\ref{mean-field-surfaces} also become softer 
along the octupole direction as one approaches $^{154}$Dy, i.e., the 
neutron octupole magic number $N=88$. In our calculations as well as in 
previous Mac-Mic  ones \cite{MM-6}, there is no static octupole 
deformation in this region. This is at variance with recent 
relativistic mean-field results \cite{microscopic-22} that predict 
octupole-deformed Dy isotopes with $N \approx 88$. However, for both $N 
\approx 88$  and  $N \approx 134$ Dy isotopes, the softness displayed 
by the Gogny-D1M MFPESs along the $Q_{30}$-direction  (see, also Fig.4 
of  Ref.~\cite{microscopic-22}), points towards the key role of 
dynamical beyond-mean-field correlations, i.e., symmetry restoration 
and/or quadrupole-octupole configuration mixing in the properties of 
the ground state and collective negative parity states in the studied 
nuclei. At this level, and at 
variance with Ref.~\cite{microscopic-22}, we conclude that the plain 
mean-field framework is not sufficient to extract conclusions 
about permanent octupole deformation effects in the 
considered nuclei. Therefore, we turn our attention to  beyond-mean-field 
correlations 
in the next Sec.~\ref{beyond-mf-results}.

\subsection{Dynamical beyond-mean-field correlations}
\label{beyond-mf-results}

In this section, we turn our attention to the impact of beyond-mean-field correlations 
in different low energy properties of the Dy isotopes considered. First,  parity projection (after variation) calculations 
are discussed in Sec.~\ref{parity-proj-results}. As shown, not only 
the MFPESs in Fig.~\ref{mean-field-surfaces}, but also the parity 
projected potential energy surfaces 
obtained for some of the considered nuclei, exhibit a rather soft 
behavior along the octupole direction with a pronounced competition 
between reflection-symmetric and reflection-asymmetric configurations. 
As a result, not only symmetry restoration but also fluctuations 
in the collective coordinates should be considered for the studied nuclei. This 
is done in Sec.~\ref{2D-GCM-results} within the framework of the 
symmetry-conserving 2D-GCM framework 
\cite{q2q3-rayner-1,q2q3-rayner-2,q2q3-rayner-3,q2q3-rayner-4}.
Since the octupole is the softest mode,
the spatial reflection symmetry is the most important 
invariance to be restored. The simultaneous restoration of 
other symmetries, such as the rotational and particle number 
symmetries \cite{Bucher-oct-2,Bernard16}, is out of the scope of the present  survey
for technical reasons such as the large number
of HO shells used and/or the number of degrees of freedom
required in the 2D-GCM ansatz.

\subsubsection{Parity symmetry restoration}
\label{parity-proj-results}

Once the intrinsic HFB states 
$ |\Phi (Q_{20},Q_{30}) \rangle$, discussed in the previous 
Sec.~\ref{mf-results}, are obtained the spatial reflection
symmetry in each of those states is restored by means of parity projection
after variation. In what follows, and for the sake of 
simplicity, we will use the notation ${\bf{Q}} =(Q_{20},Q_{30})$ for 
the pair of quadrupole and octupole deformation parameters that label each of the 
intrinsic HFB states, i.e., 
$ |\Phi (Q_{20},Q_{30}) \rangle = |\Phi ({\bf{Q}}) \rangle$. 
The projected states read
\begin{equation}
\label{PP-states}
|\Phi^{\pi} ({\bf{Q}}) \rangle = \hat{{\cal{P}}}^{\pi} |\Phi ({\bf{Q}}) \rangle
= \frac{1}{2} \Big(1 + \pi \hat{\Pi} \Big) |\Phi ({\bf{Q}}) \rangle
\end{equation}
where the projection operator $\hat{{\cal{P}}}^{\pi}$ is written in terms of the 
desired parity quantum number $\pi = \pm 1$ and the parity operator
$\hat{\Pi}$. 

\begin{figure*}
\includegraphics[width=1.00\textwidth]{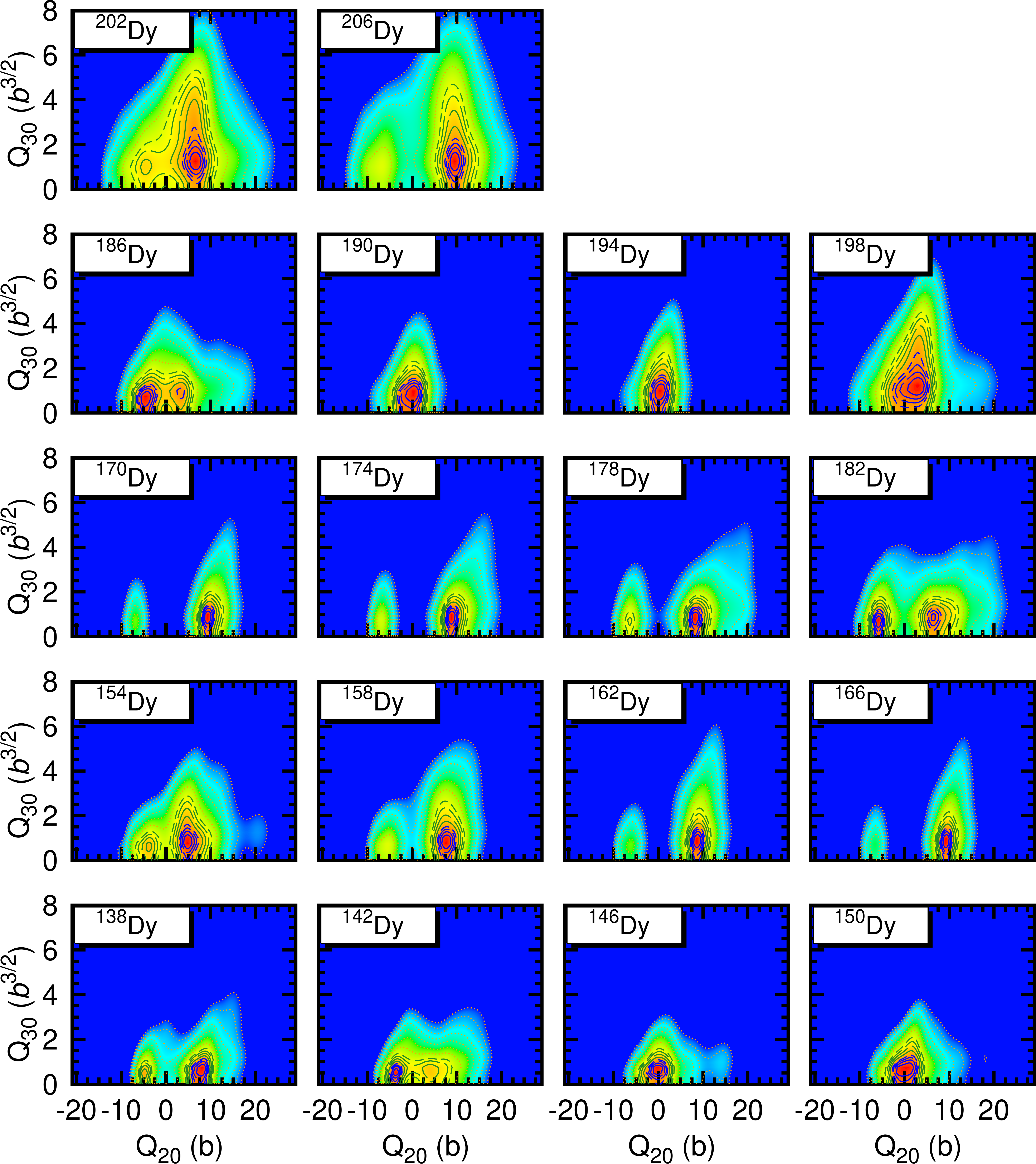}
\caption{(Color online) Positive parity ($\pi = +1$) PPPESs computed 
with the Gogny-D1M EDF  for a selected set of Dy isotopes. See, caption 
of Fig.~\ref{mean-field-surfaces}  for the contour-line patterns and 
color scale.
}
\label{PPP-surfaces} 
\end{figure*}

\begin{figure*}
\includegraphics[width=1.00\textwidth]{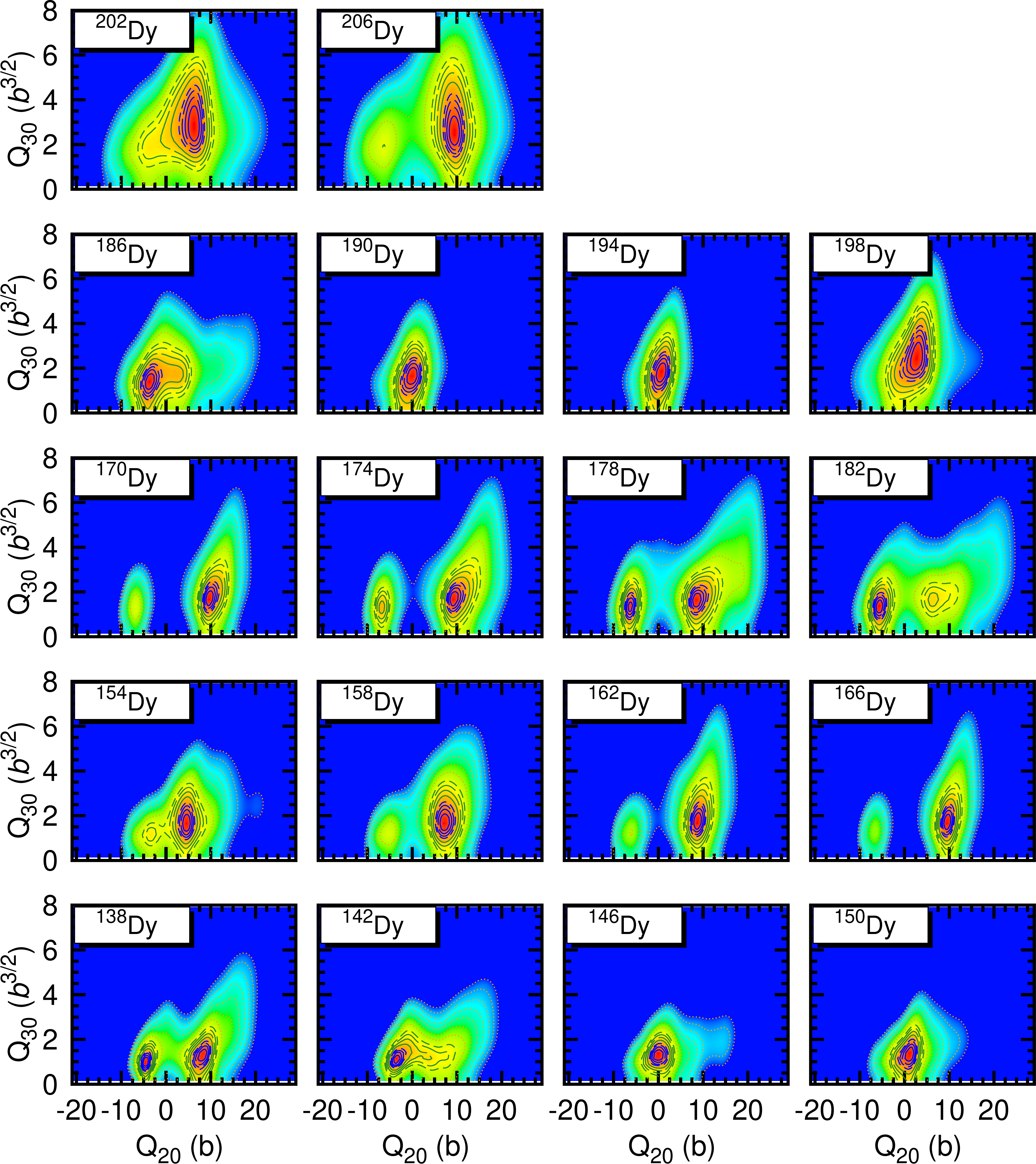}
\caption{(Color online) Negative parity ($\pi = -1$) PPPESs computed 
with the Gogny-D1M EDF  for a selected set of Dy isotopes. See, caption 
of Fig.~\ref{mean-field-surfaces}  for the contour-line patterns and 
color scale.
}
\label{NPP-surfaces} 
\end{figure*}

\begin{figure}
\includegraphics[width=0.45\textwidth]{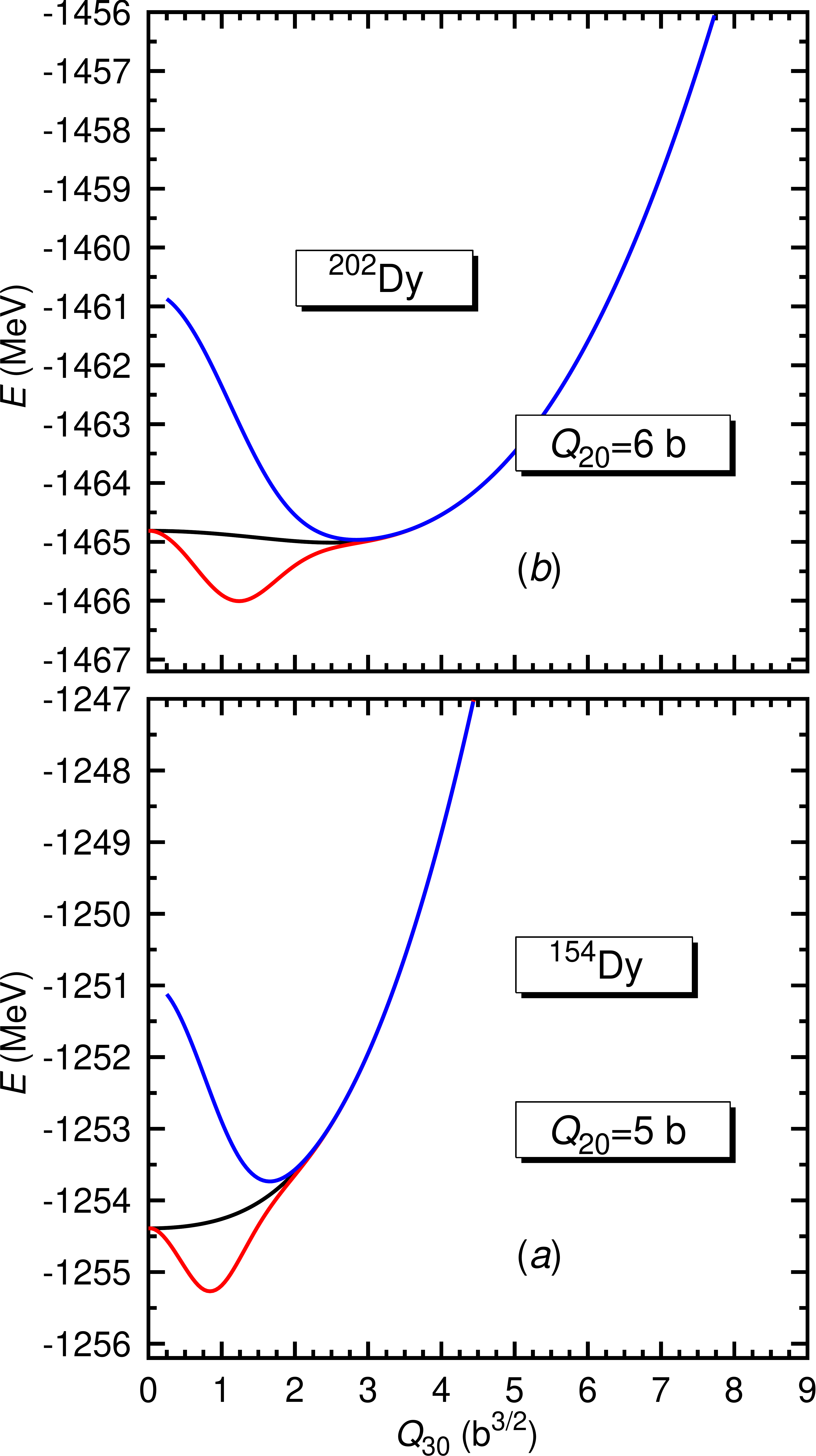}
\caption{(Color online) The $\pi = +1$ (red) and $\pi = -1$ (blue) 
parity-projected energies are depicted as functions of the octupole 
moment $Q_{30}$ for fixed values of the quadrupole moment $Q_{20}$ in 
the nuclei $^{154}$Dy and $^{202}$Dy. The corresponding HFB energies 
(black) are also included in the plots. Results have been obtained with 
the Gogny-D1M EDF.
}
\label{EXAMPLES-e-VS-q30} 
\end{figure}

In the case of the density dependent Gogny-EDF, the projected energies 
\begin{equation}
\label{PP-energies}
E_{\pi}({\bf{Q}}) 
=
\frac{\langle \Phi({\bf{Q}}) | \hat{H} {\cal{P}}^{\pi} | \Phi({\bf{Q}}) \rangle} 
{\langle \Phi ({\bf{Q}}) |{\cal{P}}^{\pi}|  \Phi ({\bf{Q}}) \rangle}
\end{equation}
associated with the parity-projected states  $|\Phi^{\pi} ({\bf{Q}}) \rangle$ 
(\ref{PP-states}), have been computed using a mixed-density prescription 
in the  density-dependent term of the EDF 
to avoid the pathologies found in the restoration of spatial symmetries
\cite{NPA-rayner-Mg,Egido-Lectures,Robledo-pathology-1,Robledo-pathology-2,Sheikh21}.
We have also introduced first-order corrections in Eq.(\ref{PP-energies}) to 
account for the fact that the parity-projected mean value of proton
and neutron numbers, usually differ from the nucleus'
proton and neutron  numbers \cite{q2q3-rayner-1,q2q3-rayner-2,q2q3-rayner-4}.
The $\pi = +1$ and $\pi = -1$ parity-projected potential energy
surfaces (PPPESs), depicted in Figs.~\ref{PPP-surfaces}  and 
\ref{NPP-surfaces} for a selected set of Dy isotopes as illustrative examples, are 
nothing else than the energies $E_{\pi}({\bf{Q}})$, as functions 
of the quadrupole $Q_{20}$ and octupole $Q_{30}$ deformations 
of the intrinsic states. As in previous studies, in  Fig.~\ref{NPP-surfaces} we have omitted the $Q_{30}=0$ 
line as the evaluation of 
$E_{\pi=-1}$ requires the non trivial task of resolving numerically a zero over zero indeterminacy. 
Fortunately, the negative parity projected energy increases rapidly as the $Q_{30}=0$ line is
approached (see, Fig.~\ref{EXAMPLES-e-VS-q30}) and  its limiting
value \cite{microscopic-8} is high enough as not to play a significant role
in the discussion of the $\pi = -1$ PPPESs \cite{q2q3-rayner-1}.

The comparison between the   PPPESs and the MFPESs in 
Fig.~\ref{mean-field-surfaces}, reveals that the quadrupole 
deformations corresponding to their absolute minima are close to each 
other. Moreover, from the comparison between the MFPESs and $\pi = +1$ 
PPPESs one realizes that, in spite of the changes in topography along 
the $Q_{30}$-direction, the latter are also rather octupole-soft and/or 
display a pronounced competition between reflection-symmetric and 
reflection-asymmetric configurations. This is illustrated in panels (a) 
and (b) of Fig.~\ref{EXAMPLES-e-VS-q30} where the $\pi = +1$ 
parity-projected energies obtained for $^{154}$Dy and $^{202}$Dy are 
plotted, as functions of $Q_{30}$, for  fixed values of the quadrupole 
moment. At the HFB level, the ground state of  $^{154}$Dy is reflection 
symmetric whereas the one of $^{202}$Dy shows a non-zero octupole 
moment. However,  for both isotopes the $\pi = +1$ parity-projected 
curves in Fig.~\ref{EXAMPLES-e-VS-q30} display a minimum with a pocket 
around $Q_{30} = 1 b^{3/2}$. In both cases, such an octupole-deformed 
minimum is less than 1.3 MeV deeper than the reflection-symmetric 
configuration indicating that, in addition to parity symmetry 
restoration, fluctuations in the collective coordinates (in particular, 
the octupole coordinate which represents the softest mode) should be 
taken into account for the studied nuclei. On the other hand, the $\pi 
= -1$ PPPESs shown in Fig.~\ref{NPP-surfaces} [see also, panels (a) and 
(b) of Fig.~\ref{EXAMPLES-e-VS-q30}] exhibit in all the cases absolute 
minima with octupole deformations larger than the ones in the MFPESs 
and  $\pi = +1$ PPPESs.

\subsubsection{Symmetry-conserving 2D-GCM quadrupole-octupole configuration mixing}
\label{2D-GCM-results}

The results discussed in  Secs.~\ref{mf-results} and 
\ref{parity-proj-results} indicate that not only parity symmetry 
restoration but also symmetry-projected quadrupole-octupole 
configuration mixing is required to disentangle the stability of 
octupole deformation effects in the studied Dy isotopes. To this end, 
we consider the following  2D-GCM superposition of HFB states $| \Phi 
({\bf{Q}})\rangle$ 
\begin{equation} \label{GCM-WF}
| {\Psi}_{\sigma}^{\pi} \rangle = \int_{{\cal{D}}} d{\bf Q} f_{\sigma}^{\pi} ({\bf Q}) | {\Phi} ({\bf Q}) \rangle
\end{equation}
where, both positive and negative octupole moments are included in the 
integration domain ${\cal{D}}$. The 2D-GCM ansatz $| 
{\Psi}_{\sigma}^{\pi} \rangle$ accounts for both reflection symmetry 
restoration and $(Q_{20},Q_{30})$-fluctuations 
\cite{q2q3-rayner-1,q2q3-rayner-2,q2q3-rayner-3,q2q3-rayner-4}. In 
Eq.(\ref{GCM-WF}) $\pi = \pm 1$  represents the parity quantum number, 
while the index $\sigma$ numbers the different GCM solutions.  

The amplitudes $f_{\sigma}^{\pi} ({\bf Q})$ should  be determined 
dynamically via the solution of the corresponding Griffin-Hill-Wheeler 
(GHW) equation \cite{rs,q2q3-rayner-1,q2q3-rayner-2,q2q3-rayner-4}, 
written in terms of non-diagonal norm ${\cal{N}}({\bf Q}, {\bf 
Q}^{'})=\langle {\Phi} ({\bf Q}) |  {\Phi} ({\bf Q}^{'}) \rangle$ and 
Hamiltonian ${\cal{H}}({\bf Q}, {\bf Q}^{'})=\langle {\Phi} ({\bf Q}) 
|\hat{H} | {\Phi} ({\bf Q}^{'}) \rangle$ overlaps. 
In the evaluation of the Hamiltonian overlap one has to pay special attention
to avoid the use of non-equivalent bases in the left and right HFB states \cite{Robledo94}.
In our case, this is accomplished by using the same oscillator lengths for
all HFB states considered in the GCM mixing \cite{Robledo22a,Robledo22b}.
For the evaluation of the density-dependent contribution of the Gogny-EDF to the Hamiltonian overlap
we have considered a mixed-density prescription in the  
density-dependent term of the EDF 
\cite{Sheikh21,q2q3-rayner-1,q2q3-rayner-2,q2q3-rayner-4}. Finally, perturbative 
first-order corrections in both the mean value of proton and neutron numbers have been
considered \cite{Sheikh21,q2q3-rayner-1,q2q3-rayner-2,q2q3-rayner-4}.

The solution of the GHW equation provides the dynamical amplitudes 
$f_{\sigma}^{\pi} ({\bf Q})$. Nevertheless, in the case of a 
non-orthogonal basis of HFB states $| {\Phi} ({\bf Q}) \rangle$, i.e., 
$\langle {\Phi} ({\bf Q}) | {\Phi} ({\bf Q}^{'}) \rangle \ne 
\delta({\bf Q}-{\bf Q}^{'})$, such amplitudes $f_{\sigma}^{\pi} ({\bf 
Q})$ cannot be assigned a quantum mechanical probabilistic 
interpretation \cite{rs}. One then introduces the collective wave 
functions  \cite{q2q3-rayner-1,rs,Sheikh21}
\begin{equation} \label{cll-wfs-HW} 
G_{\sigma}^{\pi} ({\bf Q}) =   \int d{\bf Q}^{'} {\cal
{N}}^{\frac{1}{2}}({\bf Q}, {\bf Q}^{'})  f_{\sigma}^{\pi}({\bf 
Q}^{'})
\end{equation}
written in terms of the amplitudes $f_{\sigma}^{\pi}({\bf Q})$ 
Eq.(\ref{GCM-WF}) and the operational square root 
${\cal{N}}^{\frac{1}{2}}({\bf Q}, {\bf Q}^{'})$
of the norm overlap kernel \cite{rs,Sheikh21}.

The reduced transition probabilities $B(E1,1^{-} \rightarrow 0^{+} )$ 
and $B(E3,3^{-} \rightarrow 0^{+})$ have been computed using the 
rotational model approximation for K=0 bands 
\begin{equation} \label{exp-BE1}
	B(E\lambda,\lambda^{-} \rightarrow 0^{+} ) = \frac{e^{2}}{4 \pi} 
	\Big{|} \langle \Psi_{\sigma}^{\pi=-1} | \hat{{\cal O}}_{\lambda} |\Psi_{\sigma=1}^{\pi=+1} \rangle \Big{|}^{2}
\end{equation}
where $\sigma$ corresponds to the first 2D-GCM excited negative-parity state. The 
electromagnetic transition operators  $\hat{\cal{O}}_{1}$ and $\hat{\cal{O}}_{3}$ represent the 
dipole moment operator and the proton 
component $\hat{Q}_{30, \textrm{prot}}$ of the octupole operator, respectively. 
The overlaps 
$\langle \Psi_{\sigma}^{\pi} | \hat{{\cal O}}_{\lambda} |\Psi_{\sigma^{'}}^{\pi^{'}} \rangle$
have been evaluated using the expressions given in Ref.~\cite{q2q3-rayner-1}. 

\begin{figure*}
\includegraphics[angle=-90,width=0.95\textwidth]{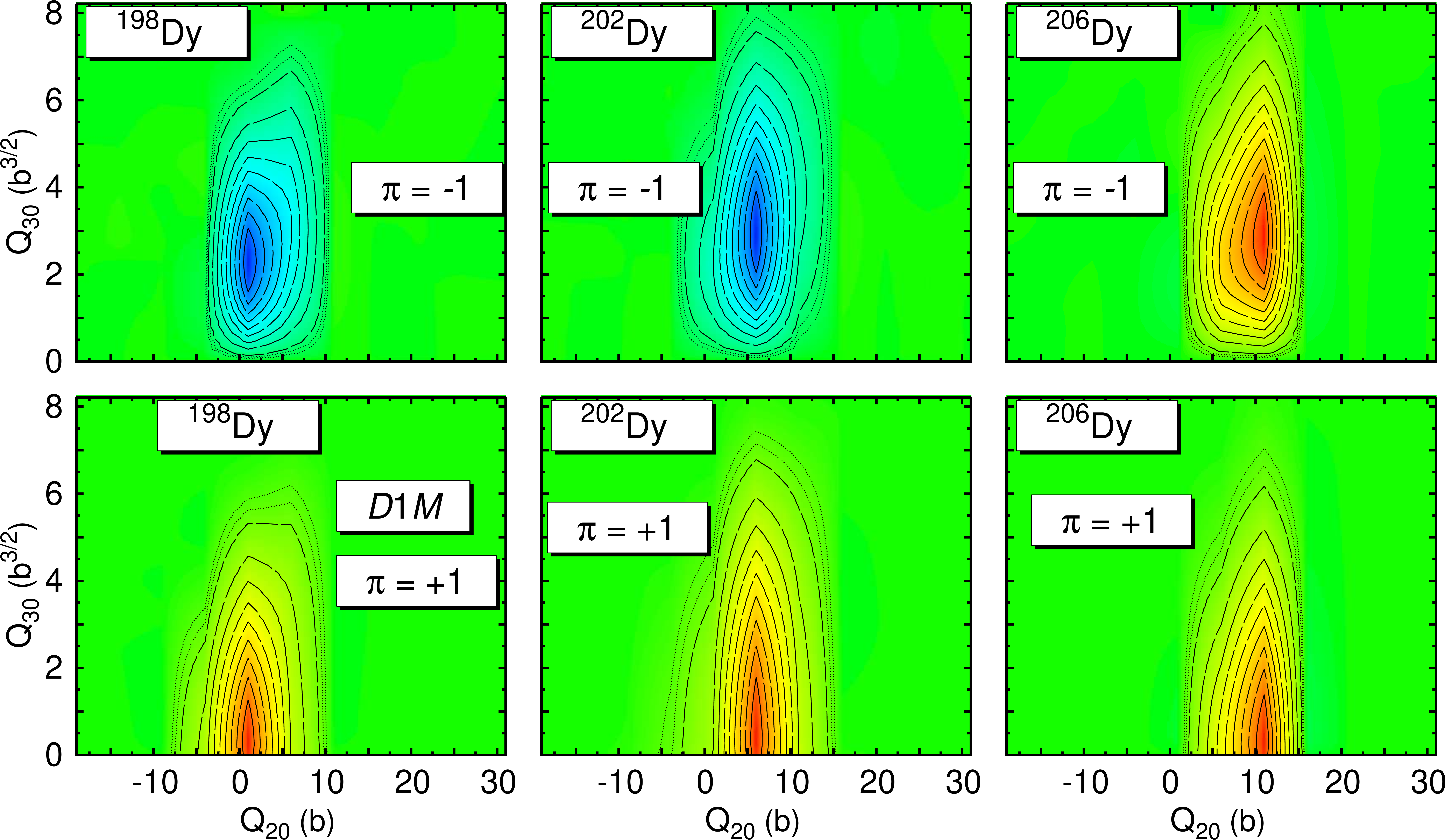}
\caption{(Color online) Collective wave functions Eq.(\ref{cll-wfs-HW})
corresponding to the ground (bottom panels) and lowest negative-parity (top panels)
states of the nuclei $^{198}$Dy, $^{202}$Dy and $^{206}$Dy. The succession of 
solid, long dashed and short dashed contour lines starts at 90$\%$ of the maximum 
value up to 10$\%$ of it. The two dotted-line contours correspond to the tail 
of the amplitude (5$\%$ and 1$\%$ of the maximum value). The color scale ranges from
red (maximum value) to green (zero). Results have been obtained with the Gogny-D1M
EDF. For more details, see the main text.
}
\label{example_CWF_POS} 
\end{figure*}

The collective wave functions Eq.(\ref{cll-wfs-HW}) corresponding to 
the ground and lowest negative-parity states of the nuclei $^{198}$Dy, 
$^{202}$Dy and $^{206}$Dy are depicted in Fig.~\ref{example_CWF_POS}, 
as illustrative examples. Similar results have been obtained for other 
Dy isotopes. Note that at the HFB level $^{198}$Dy and  $^{202}$Dy 
($^{206}$Dy) exhibit reflection-asymmetric (reflection-symmetric) 
ground states.  

The values obtained for the average quadrupole moments
\begin{equation}
	(\bar{Q}_{20})_{\sigma}^{\pi}= 
	\langle {\Psi}_{\sigma}^{\pi} | \hat{Q}_{20}| {\Psi}_{\sigma}^{\pi} \rangle.
\end{equation}
corresponding to the 2D-GCM ground states $(\bar{Q}_{20})_{\sigma=1}^{\pi=+1}$, display 
a pattern similar to the one obtained at the mean-field level [see, panel (a) of 
Fig.~\ref{summary-def-mf}]. The pattern followed by $(\bar{Q}_{20})_{\sigma=1}^{\pi=+1}$, as well
as the one followed by the average quadrupole moments corresponding 
to the first negative-parity states $(\bar{Q}_{20})_{\sigma}^{\pi=-1}$, clearly
reflect the impact of the neutron shell closures $N=82$ and $N=126$ in the 
evolution of the quadrupole properties along the considered isotopic 
chain.

The ground state collective wave functions $G_{\sigma=1}^{\pi=+1} ({\bf 
Q})$, shown in the bottom panels of Fig.~\ref{example_CWF_POS} for the 
$N \approx 134$ isotopes $^{198}$Dy, $^{202}$Dy and $^{206}$Dy exhibit  
a large spreading along the $Q_{30}$-direction. This is also the case 
for the $G_{\sigma=1}^{\pi=+1} ({\bf Q})$ amplitudes corresponding to 
$N \approx 88$ Dy isotopes. This reflects the octupole-soft character 
of the Gogny-D1M 2D-GCM ground states in the case of $N \approx 88$ and 
$N \approx 134$ Dy isotopes. However, for all the nuclei studied in 
this paper, the $G_{\sigma=1}^{\pi=+1} ({\bf Q})$ amplitudes exhibit 
peaks around $Q_{30} = 0$ pointing to an octupole-vibrational 
character. 

In order to access dynamical octupole deformation effects at a more 
quantitative level, we have computed the average octupole moment 
\cite{q2q3-rayner-1,q2q3-rayner-2,q2q3-rayner-4}
\begin{equation}
	(\bar{Q}_{30})_{\sigma}^{\pi}= 4 \int_{{\cal{D}}} d {\bf Q} d {\bf Q'} G_{\sigma}^{\pi \,*} ( {\bf Q} ) 
	 {\cal {Q}}_{30}  ({\bf Q}, {\bf Q}^{'})
G_{\sigma}^{\pi} ({\bf Q}^{'})
\end{equation}
and obtained that, for all the considered nuclei, the
ground state  $(\bar{Q}_{30})_{\sigma=1}^{\pi=+1}$ values are within the range 
$ 0.25~b^{3/2} \le (\bar{Q}_{30})_{\sigma=1}^{\pi=+1} \le 0.93~b^{3/2}$. On the one hand, this
indicates an enhanced octupolarity in their ground states via dynamical
zero-point 2D-GCM quantum fluctuations. On the other hand, even the largest 
$(\bar{Q}_{30})_{\sigma=1}^{\pi=+1}$
values obtained for $N \approx 134$ isotopes are 
less than half of the (static) HFB ground state octupole deformations. Thus, to a 
large extent, even the static octupole deformation effects predicted 
at the Gogny-HFB level around  $N = 134$ are washed out once 
symmetry-conserving quadrupole-octupole configuration mixing 
is taken into account. 

The previous results, point towards octupole-vibrational features in 
the  Dy chain, and raise questions about the conclusions extracted in 
Ref.~\cite{microscopic-22} from the results of a plain mean-field 
calculation. In this reference the existence of permanent octupole 
deformations in  $N = 88$ and $N = 134$ Dy isotopes is concluded. Let 
us stress, that results (not shown) similar to the ones already 
discussed have also been obtained in the present study with other 
parametrizations of the Gogny-EDF, such as D1S \cite{gogny-d1s} and 
D1M$^{*}$ \cite{gogny-d1mstar} (see also, the discussion below).

The collective wave functions $G_{\sigma}^{\pi=-1} ({\bf Q})$ 
corresponding to the  lowest negative-parity states of the nuclei 
$^{198}$Dy, $^{202}$Dy and $^{206}$Dy, shown in the top panels of 
Fig.~\ref{example_CWF_POS}, are odd under the exchange $Q_{30} 
\rightarrow -Q_{30}$. They  reach a zero value at $Q_{30}=0$ as well as 
a maximum and a minimum, one for a positive octupole deformation and 
the other at the corresponding negative value. As a result, the 
amplitudes $G_{\sigma}^{\pi=-1} ({\bf Q})$ in 
Fig.~\ref{example_CWF_POS} always display a maximum or a minimum for 
$Q_{30} \ne 0$. For each of the studied  isotopes, the 
$(\bar{Q}_{30})_{\sigma=}^{\pi=-1}$ value is close to the octupole 
deformation corresponding to the minimum of the $\pi =-1$ PPPESs [see, 
Fig.~\ref{NPP-surfaces}].
 
\begin{figure*}
\includegraphics[angle=-90,width=0.95\textwidth]{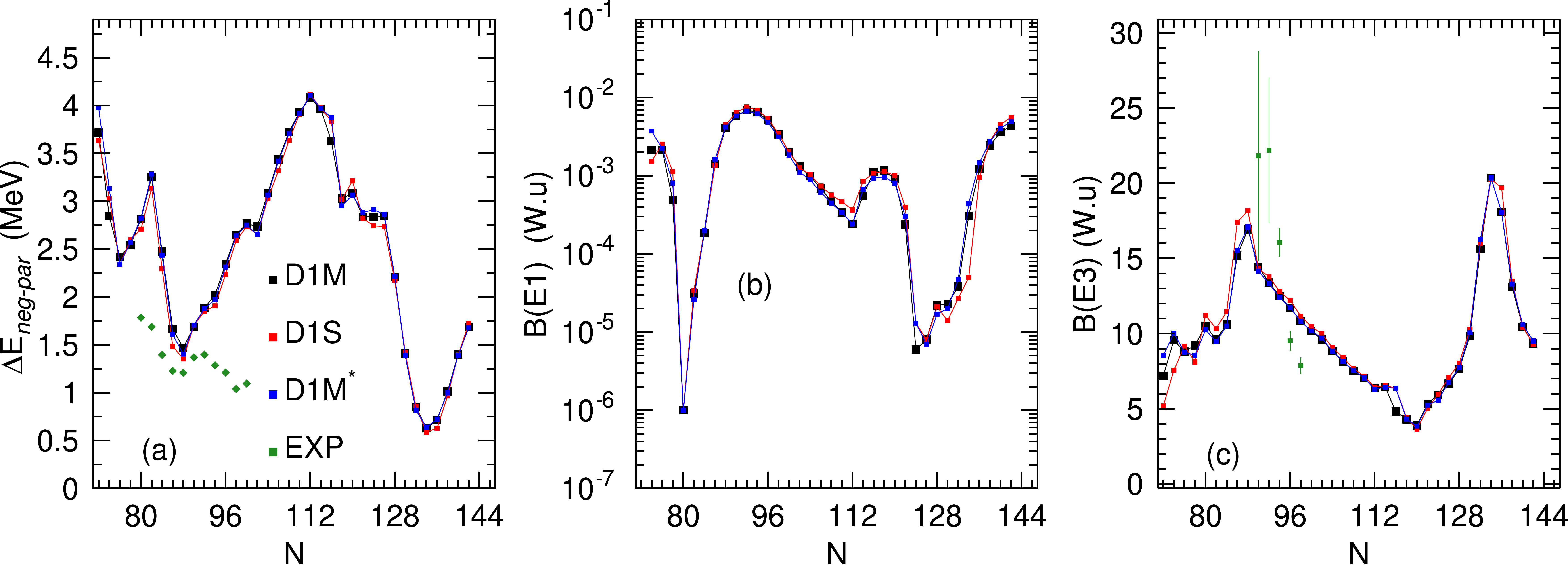}
\caption{(Color online) The 2D-GCM excitation energies of the first negative-parity 
states [panel (a)], the reduced transitions probabilities $B(E1)$ [panel (b)] and 
$B(E3)$ [panel (c)] are plotted for $^{138-208}$Dy as functions of the neutron number.
Results have been obtained with the parametrizations D1M, D1S and D1M$^{*}$ of the 
Gogny-EDF. Experimental data have been taken from Ref.~\cite{EXP-E3-Be-Kibedi}.
For more details, see the main text.
}
\label{E1BE1BE3} 
\end{figure*}

The 2D-GCM excitation  energies $\Delta E_{neg-par}$ of the first 
negative-parity states as well as the reduced transitions probabilities 
$B(E1)$ and $B(E3)$ obtained for the considered Dy isotopes, are plotted 
in panels (a), (b) and (c) of 
Fig.~\ref{E1BE1BE3}, as functions of the neutron number. Additional 
results for the parametrizations D1S and D1M$^{*}$  are also included 
in the figure. As can be seen, with minor exceptions, the results 
obtained with different parametrizations are rather similar. This 
points towards the robustness of the predicted trends with respect to 
the underlying Gogny-EDF.

As functions of the neutron number, the 2D-GCM energies $\Delta E_{neg-par}$ display
two pronounced minima, one at $N=88$ and the other at $N=134$. These Gogny 2D-GCM
results indicate that, at a dynamical beyond-mean-field level, both 
$N=88$ and  $N=134$  represent (on the average) octupole magic numbers along the 
studied isotopic chain. Let us stress that no static 
octupole deformation is obtained for $N \approx 88$ isotopes at the 
mean-field level. Moreover, the ground state collective wave functions 
for those isotopes are peaked around $Q_{30} = 0$. These results suggest that 
in the case of Dy isotopes, the octupole collectivity around 
$N = 88$ is more 
vibrational-like in character than suggested in Ref.~\cite{microscopic-22}
on the base of plain mean-field calculations. Static octupole deformations 
have been obtained at the Gogny-HFB level for $N \approx 134$ isotopes. However, as
already mentioned, their  ground state collective wave functions are also
peaked around $Q_{30} = 0$, while the corresponding  mean-field deformation effects are 
reduced to more than half once 2D-GCM zero-point fluctuations are included. This
suggests that the prominent minimum observed in panel (a) of the 
figure at $N = 134$, should also be associated with a vibrational character of the excitation instead of 
permanent octupole deformation effects.
Regarding the comparison with the still scarce 
data \cite{EXP-E3-Be-Kibedi}, the predicted $\Delta E_{neg-par}$ values reproduce 
reasonably well the experimental trend in the immediate neighborhood of 
$N=88$, while they overestimate considerably the available experimental values  
as one moves away from this neutron number.

The $B(E1)$ strengths shown in panel (b) of the same figure exhibit two 
minima, one at $N \approx 82$ and the other at $N \approx 126$. From a 
dynamical point of view, it is precisely around these neutron numbers 
where the overlap $\langle \Psi_{\sigma}^{\pi=-1} | \hat{{\cal O}}_{1} 
|\Psi_{\sigma=1}^{\pi=+1} \rangle$ (with $\hat{{\cal O}}_{1}$ being the 
dipole moment operator) reaches its minimum. Here, one should keep in 
mind, that the behavior of the $B(E1)$ strengths is not directly 
related with the one observed in the $\Delta E_{neg-par}$ energies 
and/or the $B(E3)$ reduced transition probabilities (see, below). In 
fact, via the strong dependence of the dipole moment on the underlying 
single-particle structure, the $B(E1)$ values might display strong 
suppression for some specific neutron numbers  
\cite{microscopic-8,q2q3-rayner-1,q2q3-rayner-2,q2q3-rayner-3,q2q3-rayner-4}, 
specially around neutron shell closures.

The trend observed in the predicted $B(E3)$ values correlates well with 
the one in the $\Delta E_{neg-par}$ energies, i.e., as functions of the 
neutron number the $B(E3)$ strengths exhibit two pronounced maxima at 
$N = 88$ and $N = 134$ where the $\Delta E_{neg-par}$ energies display 
two pronounced minima. The comparison with the available experimental 
data \cite{EXP-E3-Be-Kibedi} reveals that, in spite of the quantitative 
differences, the predicted $E3$-trend reproduces the increased octupole 
collectivity around $N = 88$ as well as its sudden decrease with 
increasing neutron number. We stress,  that the $E3$ collectivity 
around $N = 88$ and $N = 134$  is not the result of permanent mean-field
octupolarity around those neutron numbers, as concluded in 
Ref.~\cite{microscopic-22}, but directly reflects the key 
role played by dynamical fluctuations. In fact, via the 
structure of the corresponding collective wave functions, the 2D-GCM 
overlap $\langle \Psi_{\sigma}^{\pi=-1} | \hat{{\cal O}}_{3} 
|\Psi_{\sigma=1}^{\pi=+1} \rangle$ (with $\hat{{\cal O}}_{3}$ being the 
proton component  of the octupole operator) reflects the difference 
$|(\bar{Q}_{30})_{\sigma=1}^{\pi=+1}- 
(\bar{Q}_{30})_{\sigma}^{\pi=-1}|$ between the dynamical ground 
$(\bar{Q}_{30})_{\sigma=1}^{\pi=+1}$ and first negative-parity 
$(\bar{Q}_{30})_{\sigma}^{\pi=-1}$  state deformations, i.e., the 
larger (smaller) the difference the smaller (larger) the overlap 
$\langle \Psi_{\sigma}^{\pi=-1} | \hat{{\cal O}}_{3} 
|\Psi_{\sigma=1}^{\pi=+1} \rangle$. It is precisely the more pronounced 
(dynamical) enhancement of ground state octupolarity  (i.e., larger 
$(\bar{Q}_{30})_{\sigma=1}^{\pi=+1}$ values) obtained as one approaches 
both $N = 88$ and $N = 134$ that leads to a reduction of the difference 
$|(\bar{Q}_{30})_{\sigma=1}^{\pi=+1}- 
(\bar{Q}_{30})_{\sigma}^{\pi=-1}|$ and, therefore, to larger $B(E3)$ 
strengths as compared with the ones obtained as we move away from these two 
neutron octupole magic numbers.

\section{Summary and conclusions}
\label{conclusions}

In this paper we have carried out calculations, both at the mean-field level and beyond, to 
address the emergence and  stability of 
(static) mean-field octupole deformation effects in Dy isotopes from dripline to
dripline.To this 
end, we have resorted to the models already employed in
Refs.~\cite{q2q3-rayner-1,q2q3-rayner-2,q2q3-rayner-3,q2q3-rayner-4} in other
regions of the nuclear chart.

Contrary to recent reflection-asymmetric relativistic mean-field 
\cite{microscopic-22} but in agreement with previous Mac-Mic 
\cite{MM-6} results, at the Gogny-HFB level static octupole 
deformations have been found only for $N \approx 134$ isotopes, while 
nuclei with $N \approx 88$ exhibit reflection-symmetric ground states. 
Moreover, even in the case of nuclei with octupole deformed Gogny-D1M 
mean-field ground states (i.e., $^{198-202}$Dy), the HFB octupole 
correlation energies Eq.(\ref{oct-def-ener}) are always smaller than 
300 keV. This, as well as  the octupole-softness of the corresponding 
MFPESs, indicate that the plain mean-field framework is not sufficient 
to extract conclusions about permanent octupole deformation effects in 
Dy isotopes.

The results obtained in this paper, together with previous studies of 
the octupole dynamics in other regions of the nuclear chart 
\cite{q2q3-rayner-1,q2q3-rayner-2,q2q3-rayner-3,q2q3-rayner-4,microscopic-20,microscopic-21}, 
represent a warning to the use of the mean-field approach to extract 
conclusions on the permanent and/or vibrational nature of octupolarity 
in atomic nuclei with shallow octupole minima and/or octupole-soft 
MFPESs. Furthermore, it has been shown that the octupole-softness 
found in the MFPESs, especially around the neutron numbers $N = 88$ and 
$N = 134$, also extends to the parity-projected potential energy 
surfaces, pointing towards the key role of 2D-GCM symmetry-conserving 
configuration  mixing in the studied nuclei.

At the 2D-GCM level, zero-point quantum fluctuations associated with 
the restoration of reflection symmetry and fluctuations in the 
collective $(Q_{20},Q_{30})$  coordinates, lead to an enhanced 
octupolarity for all the considered isotopes, albeit with dynamical 
deformations less than half of the largest values obtained at the 
mean-field level. Therefore, to a large extent, the (static) mean-field 
octupole deformation effects are washed out in Dy nuclei once 2D-GCM 
fluctuations are taking into account. Our  analysis of the 2D-GCM 
collective wave functions as well as the trends of the predicted 
$\Delta E_{neg-par}$ excitation energies and $B(E3)$ strengths, 
corroborate an increased octupole collectivity in Dy isotopes  with $N 
\approx 88$ and $N \approx 134$. However, we stress that such increased  
octupolarity  is  a (dynamical)  vibrational-like effect that 
is not directly related to permanent mean-field octupole deformation in 
the considered nuclei.

The predicted $\Delta E_{neg-par}$ 
values reproduce reasonably well the available experimental data in the 
immediate neighborhood of $N=88$, while in the $B(E3)$ case the 
calculations account qualitatively for the increased octupole 
collectivity around $N = 88$ as well as its sudden decrease with 
increasing neutron number. The predicted $B(E1)$ reduced transition 
probabilities display strong 
suppression around  $N \approx 82$ and  $N \approx 126$.
Furthermore, the D1S, 
D1M$^{*}$ and D1M parameter sets provide  rather similar results, 
pointing towards the robustness of the predicted trends with respect to 
the underlying Gogny-EDF.

\begin{acknowledgments}
The work of RR was supported within the framework of the 
(distinguished researcher) Mar\'ia Zambrano Program, Ministry of Universities
and Seville University, Spain. The  
work of LMR was supported by Spanish Agencia Estatal de Investigacion (AEI)
of the Ministry of Science and Innovation under Grant No. PID2021-127890NB-I00. 
\end{acknowledgments}


\end{document}